\documentclass[12pt,a4,portrait]{article}
\usepackage[super,compress]{cite}
\usepackage{graphicx}
\usepackage[pdftex]{hyperref}
\usepackage{amsmath}
\usepackage{amssymb}
\usepackage{tocbibind}

\begin{document}

\title{\bf Invariant quantities in the multiscalar-tensor theories of gravitation}

\author{Piret Kuusk\thanks{piret.kuusk@ut.ee}, Laur J\"arv\thanks{laur.jarv@ut.ee} and Ott Vilson\thanks{ovilson@ut.ee} \\
	{\normalsize Institute of Physics, University of Tartu, Ravila 14c, Tartu 50411, Estonia}}

\maketitle

\begin{abstract}	
The aim of the current paper is to study the multiscalar-tensor theories of gravity without derivative couplings. We construct a few basic objects that are invariant under a Weyl rescaling of the metric and transform covariantly when the scalar fields are redefined. We introduce rules to construct further such objects and put forward a scheme that allows to express the results obtained either in the Einstein frame or in the Jordan frame as general ones. These so called ``translation'' rules are used to show that the parametrized post-Newtonian approximation results obtained in the aforementioned two frames indeed are the same if expressed in a general frame.
\vspace{0.2cm}
\end{abstract}


\section{Introduction}

Multiscalar-tensor gravity (MSTG) \cite{DamourEF,
Berkin:1993bt} generalizes the well known Jordan-Brans-Dicke scalar-tensor gravity (STG) by including more scalar fields non-minimally coupled to curvature. In recent years these theories have mostly attracted attention by providing models for inflation, \cite{Kaiser:2010yu,Kaiser:2012ak,Greenwood:2012aj,Kaiser:2013sna,Kallosh:2013daa,White:2012ya,White:2013ufa, Watanabe:2015eia} dark energy,\cite{Kuusk:2014sna,Rinaldi:2013lsa,Rinaldi:2014yta,Vardanyan:2015oha} and relativistic stars.\cite{Horbatsch:2015} To make a reliable use of these models the details of mathematical correspondence and physical interpretation of different MSTG conformal frames need to be understood, e.g.\ in the context of cosmological perturbations \cite{White:2012ya,White:2013ufa}, gravitational particle production \cite{Watanabe:2015eia}, or one-loop divergences \cite{Steinwachs:2011zs}.

As has been argued recently for a single field case the mathematical comparisons between the results obtained in different frames are greatly facilitated by quantities which remain invariant under the conformal Weyl rescaling of the metric and scalar field reparametrization \cite{JKSV_inv,JKSV_trans,Vilson}.
In this brief note we generalize the formalism of invariants to the multiscalar case. 

The paper is organized as follows. In Sec.\ \ref{Action_Functional_and_Transformations} we postulate an action functional for multiscalar-tensor theories of gravity, invoke the Weyl rescaling of the metric and redefinition of the scalar fields in order to study the transformation properties of the unspecified functions contained in the action. The equations of motion and the Einstein and the Jordan frame are introduced in Sec.\ \ref{Equations_of_Motion_Frames_and_Parametrizations}. Next, Sec.\ \ref{Invariants_and_metric} is devoted to the functions of the scalar fields as well as to the metric tensors for the space of scalar fields that are invariant under the local Weyl rescaling of the (spacetime) metric. We note that a spacetime Weyl rescaling induces a disformal transformation in the space of scalar fields. Based on these results, in Sec.\ \ref{Translation_Rules} we construct the so called ``translation'' rules for both the Einstein frame and the Jordan frame. These are used in Sec.\ \ref{PPN} in order to express the parametrized post-Newtonian approximation results\cite{DamourEF,Erik,Berkin:1993bt} in a generic frame.


\section{Action Functional and Transformations}
\label{Action_Functional_and_Transformations}

Let us start by postulating an action functional (generalizing Refs.\ \citen{Flanagan,JKSV_inv})
\begin{align}
\nonumber
S =& \frac{1}{2\kappa^2}\int_{V_4} \mathrm{d}^4x\sqrt{-g} \left\lbrace \mathcal{A} \,\negthickspace \left( \Phi \right)  R - \mathcal{B}_{AB} \,\negthickspace \left( \Phi \right) g^{\mu\nu} \nabla_\mu \Phi^A \nabla_\nu \Phi^B - 2\ell^{-2} \mathcal{V} \,\negthickspace \left( \Phi \right)  \right\rbrace \\
\label{MSTG_Flanagan_action}
&+ S_\mathrm{m}\left[ e^{2\alpha \left( \Phi \right)}g_{\mu\nu}, \chi \right] \,
\end{align}
describing a generic multiscalar-tensor theory of gravity without derivative couplings.\cite{DamourEF,Kaiser:2010yu,Berkin:1993bt} It contains three unspecified functions $\mathcal{A}(\Phi),\, \mathcal{V}(\Phi),\, \alpha(\Phi)$ and one invertible symmetric square matrix function $\mathcal{B}(\Phi)_{AB}$ of order $n$. Each of the three unspecified functions as well as the entries of the matrix $\mathcal{B}(\Phi)_{AB}$ in general depend on all $n$ scalar fields denoted by the set $\Phi \equiv \left\lbrace \Phi^A \right\rbrace_{A=1}^n$ as an argument of these quantities. We consider the scalar fields $\Phi^A$, the functions $\mathcal{A}(\Phi),\, \mathcal{V}(\Phi),\, \alpha(\Phi)$ and the entries of $\mathcal{B}_{AB}(\Phi)$ to be dimensionless. In order for the latter to be consistent with $c = 1$, while $\hbar$ and $G_N$ are left unspecified, we have introduced constants $\kappa^2$ having the dimension of the Newtonian gravitational constant $G_N$ and $\ell > 0$ having the dimension of length. The matter fields, collectively denoted by $\chi$, are described by the action $S_\mathrm{m}$.

The functions $\mathcal{A}(\Phi)$ and $\alpha(\Phi)$ characterize the scalar coupling to curvature and to matter, respectively. The interactions between the scalar fields are gathered into $\mathcal{V}(\Phi)$ which is often referred to as potential. The matrix $\mathcal{B}(\Phi)_{AB}$ gives the kinetic couplings of the scalar fields.

One might want to apply the local Weyl rescaling to the metric tensor $g_{\mu\nu}$ and reparametrize the scalar fields $\Phi^A$ as
\begin{subequations}
	\label{conformal_and_scalar_fields_transformations}
	\begin{align}
	\label{conformal_transformation}
	g_{\mu\nu} &= e^{2\bar{\gamma}\left( \bar{\Phi} \right)  } \bar{g}_{\mu\nu} \,, \\
	\label{scalar_fields_transformations}
	\Phi^A &= \bar{f}^A\left( \bar{\Phi} \right) \,.
	\end{align}
\end{subequations}
Often the term change of the frame is used to refer to the conformal transformation \eqref{conformal_transformation} of the metric tensor, while \eqref{scalar_fields_transformations} is dubbed the change of the parametrization. If under the transformations \eqref{conformal_and_scalar_fields_transformations} the arbitrary functions of the scalar fields, contained in the action \eqref{MSTG_Flanagan_action}, are imposed to transform as
\begin{subequations}
	\label{Flanagan_transformations}
	\begin{align}
	\label{transformation_of_A}
	\mathcal{A}(\bar{f}(\bar{\Phi})) &= e^{-2\bar{\gamma}(\bar{\Phi})}\bar{\mathcal{A}}(\bar{\Phi}) \,, \\
	\label{transformation_of_V}
	\mathcal{V}(\bar{f}(\bar{\Phi})) &= e^{-4\bar{\gamma}(\bar{\Phi})} \bar{\mathcal{V}}(\bar{\Phi}) \,, \\
	\label{transformation_of_alpha}
	\alpha(\bar{f}(\bar{\Phi})) &= \bar{\alpha}(\bar{\Phi}) - \bar{\gamma}(\bar{\Phi}) \,, \\
	\nonumber
	\mathcal{B}_{AB}(\bar{f}(\bar{\Phi})) &= e^{-2\bar{\gamma}(\bar{\Phi})} \left( \bar{f}^{C}_{\phantom{C},A} \right)^{-1} \left( \bar{f}^{D}_{\phantom{C},B}  \right)^{-1} \left\lbrace \bar{B}_{CD}(\bar{\Phi}) - 6 \bar{\gamma}_{,C} \bar{\gamma}_{,D} \bar{\mathcal{A}}(\bar{\Phi}) \right. \\
	\label{transformation_of_B}
	& \hspace{2cm} \left. + 3 \left( \bar{\gamma}_{,D} \bar{\mathcal{A}}_{,C} + \bar{\gamma}_{,C} \bar{\mathcal{A}}_{,D} \right)  \right\rbrace
	\end{align}
\end{subequations}
where $\bar{f}(\bar{\Phi}) \equiv \left\lbrace \bar{f}^{A}(\bar{\Phi}) \right\rbrace_{A=1}^{n}$, then the action functional \eqref{MSTG_Flanagan_action} is invariant up to a boundary term which we shall neglect.

Here and in the following we shall make use of the convention where the ``barred'' (``unbarred'') quantities are functions of the ``barred'' (``unbarred'') scalar fields $\left\lbrace \bar{\Phi}^A \right\rbrace$ ($\left\lbrace \Phi^A \right\rbrace$). In addition, each index $A$ written after a comma denotes a partial derivative with respect to (w.r.t.) a scalar field. If such a combination is a subscript of a ``barred'' (``unbarred'') quantity then the partial derivative is taken w.r.t. the ``barred'' (``unbarred'') scalar field, e.g.\
\begin{equation}
\bar{\gamma}_{,A} \equiv \frac{\partial \bar{\gamma}(\bar{\Phi})}{\partial \bar{\Phi}^A} \, , \qquad
\mathcal{A}_{,A} \equiv \frac{\partial \mathcal{A}(\Phi)}{\partial \Phi^A} \,.
\end{equation}
We have introduced all these conventions in order to be able to drop the arguments of the functions without generating ambiguities.

The transformation \eqref{scalar_fields_transformations} can be considered as a coordinate transformation in the $n$-dimensional space of scalar fields. Then
\begin{equation}
\bar{f}^{A}_{\phantom{A},C} \equiv \frac{\partial \bar{f}^A}{\partial \bar{\Phi}^C} \equiv \frac{\partial \Phi^A}{\partial \bar{\Phi}^C} \,.
\end{equation}
is a Jacobian matrix and $\left( \bar{f}^{C}_{\phantom{C},B} \right)^{-1} \equiv \partial \bar{\Phi}^C / \partial \Phi^B$ is its inverse.


\section{Equations of Motion, Frames and Parametrizations}
\label{Equations_of_Motion_Frames_and_Parametrizations}

Varying the action functional \eqref{MSTG_Flanagan_action} w.r.t. the metric $g^{\mu\nu}$ and w.r.t. the scalar fields $\Phi^C$ gives us the following equations of motion:
\begin{subequations}
	\label{EOM}
	\begin{align}
	\nonumber
	\mathcal{A}\left( R_{\mu\nu} - \frac{1}{2} g_{\mu\nu} R \right) &+ g_{\mu\nu}\left( \frac{1}{2}\mathcal{B}_{AB} + \mathcal{A}_{,AB} \right) g^{\sigma\rho} \nabla_\sigma \Phi^A \nabla_\rho \Phi^B - \left(\mathcal{B}_{AB} + \mathcal{A}_{,AB} \right) \nabla_\mu \Phi^A \nabla_\nu \Phi^B \\
	\label{EOM_metric}
	&+ \mathcal{A}_{,A} \left( g_{\mu\nu}\Box \Phi^A - \nabla_\mu\nabla_\nu\Phi^A \right) + \ell^{-2} g_{\mu\nu} \mathcal{V} - \kappa^2 T_{\mu\nu} = 0 \,,
	\end{align}
	\begin{align}
	\nonumber
	\mathcal{A}_{,C} R &+ 2\mathcal{B}_{AC} \Box \Phi^A + \left( 2 \mathcal{B}_{BC,A} - \mathcal{B}_{AB,C} \right) g^{\mu\nu} \nabla_\mu \Phi^A \nabla_\nu \Phi^B - 2\ell^{-2} \mathcal{V}_{,C} \\
	\label{EOM_scalar_field} &+ 2\kappa^2 \alpha_{,C} T = 0 \,.
	\end{align}
\end{subequations}

Often in the literature some of the unspecified functions contained in the action \eqref{MSTG_Flanagan_action} are given a fixed functional form, e.g.\ in order to have a more straightforward physical interpretation. Let us recall the two setups that are often used.
\begin{itemize}
	\item For the Einstein frame as used in Ref.\ \citen{DamourEF} let us denote the metric tensor as $g^{\mathfrak{E}}_{\mu\nu}$ and specify the scalar functions as
	\begin{equation}
	\label{Einstein_frame}
	\mathcal{A} \equiv 1 \equiv \mathcal{A}_\mathfrak{E} \,,\qquad \mathcal{B}_{AB} \equiv 2 \mathcal{B}^{\mathfrak{E}}_{AB} \, , \qquad \mathcal{V} \equiv \mathcal{V}_\mathfrak{E} \, , \qquad \alpha \equiv \alpha_\mathfrak{E}\,.
	\end{equation}
	A closer look to the equations of motion \eqref{EOM} reveals that if $\mathcal{A}\equiv 1$ then Eq.\ \eqref{EOM_metric} does not contain the second derivatives of the scalar fields $\Phi^A$ and hence purely describes the propagation of the metric tensor $g^{\mathfrak{E}}_{\mu\nu}$. Analogously Eq.\ \eqref{EOM_scalar_field} does not contain the second derivatives of the metric tensor $g^{\mathfrak{E}}_{\mu\nu}$ and hence describes the propagation of the scalar fields $\Phi^A$. One can further separate the scalar fields by multiplying Eq. \eqref{EOM_scalar_field} with the inverse matrix $\mathcal{B}^{\mathfrak{E}\,BA}$ where $\mathcal{B}^{\mathfrak{E}\,BA} \mathcal{B}^{\mathfrak{E}}_{AC} \equiv \delta^{B}_{C}$. It is said that the equations are fully debraided.\cite{Bettoni_Zumalacarregui}\\
	\item For the Jordan frame in the Brans-Dicke-Bergmann-Wagoner (BDBW) parametrization as used in Refs. \citen{Erik,Kuusk:2014sna} let us denote the metric tensor as $g^{\mathfrak{J}}_{\mu\nu}$ and distinguish one scalar field $\Psi$ while the others are denoted as $\bar{\Phi}^a$, $a,b = 1 \ldots n-1$, where
	\begin{subequations}
	\label{Jordan_frame}
	\begin{align}
	\mathcal{A} &\equiv \Psi \equiv \bar{\mathcal{A}}_\mathfrak{J} \,, \qquad \mathcal{V} \equiv \bar{\mathcal{V}}_\mathfrak{J} \, , \qquad \alpha \equiv 0 = \bar{\alpha}_\mathfrak{J} \,, \\
	\label{conditions_on_B}
	\mathcal{B}_{ab} &\equiv \bar{\mathcal{B}}^{\mathfrak{J}}_{ab} \, , \quad \mathcal{B}_{na} \equiv 0 = \bar{\mathcal{B}}^{\mathfrak{J}}_{na} \, , \quad \mathcal{B}_{nn} \equiv \bar{\mathcal{B}}_{nn} \equiv \frac{ \omega( \bar{\Phi}^1,\ldots,\bar{\Phi}^{n-1},\Psi ) }{ \Psi } \,.
	\end{align}
\end{subequations}
The quantities in the Jordan frame are ``barred'' for the sake of notational consistency and the reason will be made clear in Subsec. \ref{subsec:Jordan_frame}. In this frame the action $S_\mathrm{m}$ for the matter fields functionally depends on the geometrical metric $g^{\mathfrak{J}}_{\mu\nu}$ and hence freely falling particles follow the geodesics of the metric $g^{\mathfrak{J}}_{\mu\nu}$.
\end{itemize}


\section{Invariants and the Metric for the Space of Scalar Fields}\label{Invariants_and_metric}

Just as in the case of one scalar field,\cite{JKSV_inv} a closer inspection of the transformation rules \eqref{Flanagan_transformations} allows us to write out two quantities that are invariant under a Weyl rescaling of the metric \eqref{conformal_transformation} and transform as scalar functions under the scalar fields reparametrization \eqref{scalar_fields_transformations}
\begin{equation}
\label{invariants}
\mathcal{I}_1(\Phi) \equiv \frac{e^{2\alpha(\Phi)}}{\mathcal{A}(\Phi)} \, , \qquad \mathcal{I}_2(\Phi) \equiv \frac{\mathcal{V}(\Phi)}{ \left( \mathcal{A}(\Phi) \right)^2 } \,.
\end{equation}
We shall call them invariants. Also an arbitrary function of these, e.g.\
\begin{equation}
\mathcal{I}_4 \equiv \frac{\mathcal{I}_2}{\mathcal{I}_1^2} = e^{-4\alpha}\mathcal{V}
\end{equation}
is an invariant.\cite{JKSV_inv} Note that these quantities are also invariants of a spacetime point. In comparison with the one scalar field case,\cite{JKSV_inv} the third invariant $\mathcal{I}_3$ and the other two rules for constructing further invariants do not generalize so straightforwardly to the case of $n$ scalar fields. To address this issue, a few preluding remarks about the metric of the space of scalar fields are in order.

One could take $\mathcal{B}_{AB}$ to be the metric of the space of scalar fields and indeed if only the scalar fields reparametrizations \eqref{scalar_fields_transformations} are considered then $\mathcal{B}_{AB}$ transforms as a second order covariant tensor. However, if also the local Weyl rescaling of the (spacetime) metric tensor is utilized, then $\mathcal{B}_{AB}$ gains additive terms. Our aim is to construct quantities that are invariant under a Weyl rescaling \eqref{conformal_transformation} and transform covariantly under scalar fields reparametrizations \eqref{scalar_fields_transformations}. Thus, we introduce the metric of the space of scalar fields and its transformation rule as 
\begin{equation}
\label{definition_of_F}
\mathcal{F}_{AB} \equiv \frac{ 2\mathcal{A} \mathcal{B}_{AB} + 3\mathcal{A}_{,A} \mathcal{A}_{,B} }{4\mathcal{A}^2} \, ,  \qquad \mathcal{F}_{AB} = \left( \bar{f}^{C}_{\phantom{C},A} \right)^{-1} \left( \bar{f}^{D}_{\phantom{D},B} \right)^{-1} \bar{\mathcal{F}}_{CD} \,.
\end{equation}
This allows us to generalize the third invariant $\mathcal{I}_3$ as an indefinite integral
\begin{equation}
\label{I_3}
\mathcal{I}_3(\Phi) \equiv \int \sqrt{\det|\mathcal{F}_{AB}|} \,\mathrm{d}\Phi^1\wedge\ldots\wedge \mathrm{d}\Phi^n \,.
\end{equation}

We assume $\mathcal{F}_{AB}$ to be an invertible matrix and denote its inverse as $\mathcal{F}^{BC}$. Introducing a covariant derivative in the space of scalar fields via the metric $\mathcal{F}_{AB}$ guarantees that the obtained differential operator is invariant under the Weyl rescaling \eqref{conformal_transformation} of the spacetime metric. Note that $\mathcal{F}^{AB}$ can be used to contract indexes as e.g.\
\begin{equation}
\label{I_5}
\mathcal{I}_5 \equiv \frac{1}{4} \mathcal{F}^{AB} \left(\ln \mathcal{I}_1 \right)_{,A} \left(\ln \mathcal{I}_1 \right)_{,B}
\end{equation}
and thereby allows us to introduce further invariants.

It is possible to define other objects that transform exactly as $\mathcal{F}_{AB}$. Namely, one could consider an invariant (Eq.\ \eqref{invariants} etc.) as a scalar function defined on the space of scalar fields and invoke a (special) disformal transformation\cite{Bekenstein} of $\mathcal{F}_{AB}$
\begin{subequations} 
\label{G}
\begin{align}
\nonumber
\mathcal{G}_{AB} \equiv& \frac{2}{\mathcal{I}_1} \mathcal{F}_{AB} - \frac{3}{2\mathcal{I}_1} \left( \ln \mathcal{I}_1 \right)_{,A}\left( \ln \mathcal{I}_1 \right)_{,B} \\
\label{definition_of_G}
=& e^{-2\alpha} \left( \mathcal{B}_{AB} - 6 \mathcal{A} \alpha_{,A}\alpha_{,B} + 3\left( \alpha_{,A} \mathcal{A}_{,B} + \alpha_{,B} \mathcal{A}_{,A} \right) \right) \,, \\
\label{definition_of_G_inverse}
\mathcal{G}^{BC} =& \frac{\mathcal{I}_1}{2}  \mathcal{F}^{BC} + \frac{ \mathcal{I}_1 }{2}  \left( 1 - 3\mathcal{I}_5 \right)^{-1} \frac{3}{4} \mathcal{F}^{ BE } \left( \ln \mathcal{I}_1 \right)_{,E} \mathcal{F}^{CF} \left( \ln \mathcal{I}_1 \right)_{,F} \,
\end{align}
\end{subequations}
where the inverse is calculated by making use of the knowledge about disformal transformations (cf. appendix A in Ref.\ \citen{disformal}). Therefore the matrix $\mathcal{G}_{AB}$ also fulfils the requirements of the metric of the space of scalar fields, and can be invoked to construct invariants analogously to Eqs.\ \eqref{I_3} and \eqref{I_5}.

Let us take a closer look to the relation between metrics \eqref{definition_of_F}, \eqref{definition_of_G} of the space of scalar fields and $\mathcal{B}_{AB}$. If one chooses to work within the Einstein frame defined by Eq.\ \eqref{Einstein_frame}, then $\left. \mathcal{F}_{AB} \right|_{\mathfrak{E}} = \mathcal{B}^{\mathfrak{E}}_{AB}$. If instead the Jordan frame, defined by Eq.\ \eqref{Jordan_frame}, is considered then $\left. \mathcal{G}_{AB} \right|_{\mathfrak{J}} = \mathcal{B}^{\mathfrak{J}}_{AB}$. In this sense, we see that a Weyl rescaling (conformal transformation) in the spacetime introduces a disformal transformation in the space of scalar fields (cf.\ also Ref.\ \citen{Kaiser}). However this relation is somewhat formal because $\mathcal{I}_1$ does not have a dynamics of its own.


\section{Translation Rules}
\label{Translation_Rules}

For one scalar field case a prescription was developed\cite{JKSV_inv,Vilson} how to easily ``translate'' the results obtained in a particular frame and parametrization to the general one. The idea is to write the action in terms of invariant quantities in the form resembling a particular frame and parametrization and read off the correspondences. In the current paper we generalize this approach to the multiscalar field case.


\subsection{Einstein frame}

Let us consider the Einstein frame setup \eqref{Einstein_frame} used by Damour and Esposito-Far\`{e}se.\cite{DamourEF} We start by defining a spacetime metric
\begin{equation}
\label{invariant_metric_Einstein_frame}
\hat{g}^{(\mathfrak{E})}_{\mu\nu} \equiv \mathcal{A}g_{\mu\nu} \,
\end{equation}
that, in the Einstein frame ($\mathcal{A} = 1$), coincides with the metric $g^{\mathfrak{E}}_{\mu\nu}$. Note that due to suitable transformation properties of $\mathcal{A}$, given by Eq.\ \eqref{transformation_of_A}, the metric $\hat{g}^{(\mathfrak{E})}_{\mu\nu}$ does not transform under the local Weyl rescaling \eqref{conformal_transformation}. Because of that we shall use the term invariant metric to refer to $\hat{g}^{(\mathfrak{E})}_{\mu\nu}$ and other metric tensors having the same transformation properties. Expressing the action \eqref{MSTG_Flanagan_action} in terms of the new dynamical metric $\hat{g}^{(\mathfrak{E})}_{\mu\nu}$, while neglecting the boundary term, we get
\begin{align}
\nonumber
S =& \frac{1}{2\kappa^2}\int_{V_4} \mathrm{d}^4x\sqrt{-\hat{g}^{(\mathfrak{E})} } \left\lbrace \hat{R}^{(\mathfrak{E})} - 2\mathcal{F}_{AB} \,\hat{g}^{(\mathfrak{E})\mu\nu} \hat{\nabla}^{(\mathfrak{E})}_\mu \Phi^A \hat{\nabla}^{(\mathfrak{E})}_\nu \Phi^B - 2\ell^{-2} \mathcal{I}_2 \right\rbrace \\
\label{MSTG_Flanagan_action_Einstein_frame}
&+ S_\mathrm{m}\left[ \mathcal{I}_1 \hat{g}^{(\mathfrak{E})}_{\mu\nu}, \chi \right] \,.
\end{align}
The obtained action has preserved all degrees of freedom and hence is as general as action \eqref{MSTG_Flanagan_action}. However, if one fixes the frame to be the Einstein frame (Eq.\ \eqref{Einstein_frame}) then we have the following mapping
\begin{equation}
\label{mapping_in_Einstein_frame}
\begin{tabular}{rcl|rcl}
$\hat{g}^{(\mathfrak{E})}_{\mu\nu} $ & $\mapsto$ & $g^{\mathfrak{E}}_{\mu\nu}$ \qquad & \qquad $1$ & $\mapsto$ & $1 \equiv \mathcal{A}_\mathfrak{E}$ \, \\
$\sqrt{-\hat{g}^{(\mathfrak{E})}}$ & $\mapsto$ & $\sqrt{-g^{\mathfrak{E}}}$ \qquad &\qquad $\mathcal{F}_{AB} \equiv  \frac{ 2\mathcal{A}\mathcal{B}_{AB} + 3 \mathcal{A}_{,A} \mathcal{A}_{,B} }{4\mathcal{A}^2} $ & $\mapsto$ & $ \mathcal{B}^{\mathfrak{E}}_{AB}$ \, \\
$\hat{R}^{(\mathfrak{E})}$  & $\mapsto$ & $R_{\mathfrak{E}}$ \qquad & \qquad $\mathcal{I}_2\equiv \frac{\mathcal{V}}{\mathcal{A}^2}$ & $\mapsto$ & $\mathcal{V}_\mathfrak{E}$ \, \\ \rule{0ex}{3.2ex}
$\hat{\nabla}^{(\mathfrak{E})}_\mu$ &  $\mapsto$ & $\nabla^{\mathfrak{E}}_\mu$ \qquad & \qquad $\frac{1}{2} \ln \mathcal{I}_1 \equiv \frac{1}{2} \ln \left( \frac{e^{2\alpha}}{\mathcal{A}} \right)$ & $\mapsto$ & $\alpha_\mathfrak{E}$ \,
\end{tabular}
\end{equation}
where $(\mathfrak{E})$ as a superscript or a subscript denotes that the quantity under consideration is calculated via the invariant metric defined by Eq.\ \eqref{invariant_metric_Einstein_frame} and $\mathfrak{E}$ (without parenthesis) denotes that the quantity is expressed in the Einstein frame.\cite{Vilson}

When one wants to express an invariant quantity calculated in the Einstein frame as a general result then one has to use the mapping \eqref{mapping_in_Einstein_frame} backwards and evaluate everything in terms of $g_{\mu\nu}$, $\mathcal{A}$, $\mathcal{V}$, $\alpha$ and $\mathcal{B}_{AB}$. Note that as no scalar fields redefinition is used for obtaining action \eqref{MSTG_Flanagan_action_Einstein_frame} also the derivative $\frac{\partial \phantom{\Phi^A} }{\partial \Phi^A}$ is mapped to itself in both directions.


\subsection{Jordan frame}
\label{subsec:Jordan_frame}

Now let us consider the Jordan frame in the Brans-Dicke-Bergmann-Wagoner type parametrization \eqref{Jordan_frame}.\cite{Erik,Kuusk:2014sna} Analogously to the previous case, we define an invariant metric
\begin{equation}
\label{invariant_metric_in_Jordan_frame}
\hat{g}^{(\mathfrak{J})}_{\mu\nu} = e^{2\alpha}g_{\mu\nu} \,.
\end{equation}
Expressing the action functional \eqref{MSTG_Flanagan_action} in terms of $\hat{g}^{(\mathfrak{J})}_{\mu\nu}$ we get
\begin{align}
\nonumber
S =& \frac{1}{2\kappa^2}\int_{V_4} \mathrm{d}^4x\sqrt{ -\hat{g}^{(\mathfrak{J})} } \left\lbrace \frac{1}{\mathcal{I}_1} \hat{R}^{(\mathfrak{J})} - \mathcal{G}_{AB} \,\hat{g}^{(\mathfrak{J})\mu\nu} \hat{\nabla}^{(\mathfrak{J})}_\mu \Phi^A \hat{\nabla}^{(\mathfrak{J})}_\nu \Phi^B - 2\ell^{-2} \mathcal{I}_4 \right\rbrace \\
\label{MSTG_Flanagan_action_Jordan_frame}
 & + S_\mathrm{m}\left[ \hat{g}^{(\mathfrak{J})}_{\mu\nu}, \chi \right] \,.
\end{align}
As before, we have neglected the boundary term. The action functional \eqref{MSTG_Flanagan_action_Jordan_frame} could be used to read off the ``translation rules'' for Jordan frame in a generic parametrization. However, in the current paper we also consider the case where the parametrization is chosen to be the BDBW parametrization as given by Eq.\ \eqref{Jordan_frame}. 

In Refs.\ \citen{Erik,Kuusk:2014sna} one scalar field has been made distinct by defining $\bar{\mathcal{A}}_\mathfrak{J} = \Psi = \left. \frac{1}{\mathcal{I}_1} \right|_{\mathfrak{J}}$ which multiplies the Ricci scalar. Following that line of thought we redefine the scalar fields $\left\lbrace \Phi^A \right\rbrace \to \left\lbrace \bar{\Phi}^1,\ldots,\bar{\Phi}^{n-1},1/\mathcal{I}_1  \right\rbrace$ in order to distinguish $1/\mathcal{I}_1$ as a scalar field that has vanishing kinetic coupling to the other scalar fields, thereby mimicking conditions \eqref{conditions_on_B}. Therefore for the latter the condition
\begin{equation}
\label{condition_on_G}
\bar{\mathcal{G}}^{an} \equiv \frac{\partial \bar{\Phi}^a}{\partial \Phi^A} \mathcal{G}^{AB} \left( \frac{1}{\mathcal{I}_1} \right)_{,B} = 0 \, , \qquad a=1\,, \ldots \,, n-1
\end{equation}
must hold. Note that this is just a transformation of $\mathcal{G}^{AB}$ under a change of coordinates in the space of scalar fields. In the same spirit
\begin{equation}
\label{G_nn}
\bar{\mathcal{G}}^{nn} = \mathcal{G}^{AB} \left( \frac{1}{ \mathcal{I}_1 } \right)_{,A} \left( \frac{1}{ \mathcal{I}_1 } \right)_{,B} =  \frac{ 2 \mathcal{I}_5}{ \mathcal{I}_1 \left( 1 - 3 \mathcal{I}_5 \right)} = \left( \bar{\mathcal{G}}_{nn} \right)^{-1} \,
\end{equation}
where the last equality follows from the condition \eqref{condition_on_G}. We also made use of Eqs.\ \eqref{G} to write the expression in terms of $\mathcal{F}^{AB}$ hidden in $\mathcal{I}_5$. Hence we see that the kinetic term for $1/\mathcal{I}_1$ is an invariant by itself,\cite{Vilson}
\begin{equation}
\bar{\mathcal{G}}_{nn} \hat{g}^{(\mathfrak{J})\mu\nu}\hat{\nabla}^{(\mathfrak{J})}_\mu \frac{1}{\mathcal{I}_1} \hat{\nabla}^{(\mathfrak{J})}_\nu \frac{1}{\mathcal{I}_1} = \frac{ \mathcal{I}_1 \left( 1 - 3 \mathcal{I}_5 \right) }{ 2 \mathcal{I}_5} \hat{g}^{(\mathfrak{J})\mu\nu}\hat{\nabla}^{(\mathfrak{J})}_\mu \frac{1}{\mathcal{I}_1} \hat{\nabla}^{(\mathfrak{J})}_\nu \frac{1}{\mathcal{I}_1} \,.
\end{equation}
In addition, due to the condition \eqref{condition_on_G} and to the result \eqref{G_nn} it holds that
\begin{equation}
\bar{\mathcal{G}}^{nn} \frac{\partial \Phi^A}{ \partial \frac{1}{ \mathcal{I}_1 } } +  \bar{\mathcal{G}}^{ a n } \frac{\partial \Phi^A}{ \partial \bar{\Phi}^{a} } = \frac{ 2 \mathcal{I}_5}{ \mathcal{I}_1 \left( 1 - 3 \mathcal{I}_5 \right)} \frac{\partial \Phi^A}{ \partial \frac{1}{\mathcal{I}_1 } } = \mathcal{G}^{AB} \left( \frac{ 1 }{ \mathcal{I}_1 } \right)_{,B} \,.
\end{equation}
Thereby we can introduce a differential operator
\begin{equation}
\label{invariant_derivative}
\frac{\partial \phantom{\Psi} }{ \partial \frac{1}{\mathcal{I}_1} } = \frac{ \partial \Phi^{A} }{\partial \frac{1}{\mathcal{I}_1 }} \frac{\partial \phantom{\Phi^A} }{ \partial \Phi^A } = \bar{\mathcal{G}}_{nn} \mathcal{G}^{AB} \left( \frac{ 1 }{ \mathcal{I}_1 } \right)_{,B} \frac{\partial \phantom{\Phi^A} }{ \partial \Phi^A }
= -\frac{ \mathcal{I}_1 }{ 4 \mathcal{I}_5 } \mathcal{F}^{AB} \left( \ln \mathcal{I}_1 \right)_{,B} \frac{\partial \phantom{\Phi^A} }{ \partial \Phi^A} \,
\end{equation}
that gives an invariant if acted upon an invariant.

The ``translation'' rules can be read out from the mapping 
\begin{equation}
\label{mapping_in_Jordan_frame}
\begin{tabular}{rcl|rcl}
$\hat{g}^{(\mathfrak{J})}_{\mu\nu}$ & $\mapsto$ & $g^{\mathfrak{J}}_{\mu\nu}$ \qquad  & \qquad $\frac{1}{\mathcal{I}_1}$ & $\mapsto$ & $\Psi \equiv \bar{\mathcal{A}}_\mathfrak{J}$ \, \\
\rule{0ex}{3ex}
$\sqrt{-\hat{g}^{(\mathfrak{J})} }$ & $\mapsto$ & $\sqrt{-g^{\mathfrak{J}} }$ \qquad& \qquad $ \frac{ \mathcal{I}_1 \left( 1 - 3\mathcal{I}_5 \right) }{ 2 \mathcal{I}_5} $ & $\mapsto$ & $\frac{\omega}{\Psi} \equiv \bar{\mathcal{B}}^{\mathfrak{J}}_{nn}$ \, \\
$\hat{R}^{(\mathfrak{J})}$ & $\mapsto$ & $R_\mathfrak{J}$ \qquad & \qquad $\mathcal{I}_4 \equiv e^{-4\alpha} \mathcal{V}$ & $\mapsto$ & $\bar{\mathcal{V}}_\mathfrak{J}$ \, \\
$\hat{\nabla}^{(\mathfrak{J})}_{\mu}$ & $\mapsto$ & $\nabla^{\mathfrak{J}}_{\mu}$ \qquad & \qquad $0$ & $\mapsto$ & $0 = \bar{\alpha}_\mathfrak{J}$ \, \\
$\hat{R}^{(\mathfrak{J})}_{\mu\nu}$ & $\mapsto$ & $R_{\mathfrak{J}\,\mu\nu}$ \qquad & \qquad $-\frac{ \mathcal{I}_1 }{ 4 \mathcal{I}_5 }  \left( \ln \mathcal{I}_1 \right)^{ ,A } \frac{\partial \phantom{\Phi^A} }{ \partial \Phi^A}$ & $\mapsto$ & $\frac{\partial \phantom{\Psi} }{ \partial \Psi}$ \,
\end{tabular}
\end{equation}
where, analogously to the previous case the superscript $(\mathfrak{J})$ denotes a quantity calculated via the invariant metric \eqref{invariant_metric_in_Jordan_frame} and super or subscript $\mathfrak{J}$ indicates that the quantity under consideration is evaluated in the Jordan frame. The mapping for $\frac{\omega}{\Psi}$ follows from Eq.\ \eqref{G_nn}, while $\frac{\partial \phantom{\Psi} }{ \partial \Psi }$ is Eq.\ \eqref{invariant_derivative} where the indexes are raised with $\mathcal{F}^{AB}$. Similarly to the Einstein frame, if one wants to ``translate'' invariant quantities then one has to invoke the mapping \eqref{mapping_in_Jordan_frame} backwards. Note that the rules given by Eq.\ \eqref{mapping_in_Jordan_frame} are not complete but they are sufficient for showing how the formalism works as is done in the next section.


\section{Parametrized Post-Newtonian Approximation}
\label{PPN}

Each theory must be confronted with experiments. For metric gravity theories a prescription named the parametrized post-Newtonian approximation (PPN) has been constructed in order to be able to test the viability of a theory via experiments carried out in the solar system.

In the current paper we do not calculate the PPN parameters but rather show that the results obtained in different frames generalize to the same invariant and hence are frame-independent. We start by writing out the results from Ref.\ \citen{DamourEF} where the Einstein frame (without potential) was considered:
\begin{subequations}
\label{PPN_in_Einstein_frame}
\begin{align}
G_{eff} &\equiv \frac{ \kappa^2 }{8\pi} e^{2\alpha_{\mathfrak{E}}}  \left( 1 + \mathcal{B}^{\mathfrak{E}\,AB} \left( \alpha_\mathfrak{E} \right)_{,A} \left( \alpha_\mathfrak{E} \right)_{,B} \right) \,, \\
\gamma - 1 &\equiv -2 \left( \frac{\mathcal{B}^{\mathfrak{E}\,AB} \left( \alpha_{\mathfrak{E}} \right)_{,A} \left( \alpha_{\mathfrak{E}} \right)_{,B} }{ 1 + \mathcal{B}^{\mathfrak{E}\,AB} \left( \alpha_{\mathfrak{E}} \right)_{,A} \left( \alpha_{\mathfrak{E}} \right)_{,B} } \right) \,, \\
\beta - 1 &\equiv \frac{\mathcal{B}^{\mathfrak{E}\,AC} \left( \alpha_{\mathfrak{E}} \right)_{,C} \left( \left( \alpha_{\mathfrak{E}} \right)_{,AB} - \Gamma^{F}_{AB} \left( \alpha_{\mathfrak{E}} \right)_{,F} \right)  \mathcal{B}^{\mathfrak{E}\,BD} \left( \alpha_{\mathfrak{E}} \right)_{,D} }{2 \left( 1 + \mathcal{B}^{\mathfrak{E}\,AB} \left( \alpha_{\mathfrak{E}} \right)_{,A} \left( \alpha_{\mathfrak{E}} \right)_{,B} \right)^2} \,
\end{align}
\end{subequations}
where $\Gamma^{F}_{AB}$ are the Christoffel symbols for $\mathcal{B}^{\mathfrak{E}}_{AB}$.

Second, we write out the results from Ref.\ \citen{Erik} obtained in the Jordan frame (without potential):
\begin{subequations}
\label{PPN_in_Jordan_frame}
\begin{align}
G_{eff} &\equiv \frac{ \kappa^2 }{8\pi} \frac{1}{ \Psi } \left( 1 + \frac{1}{ 2\omega + 3 } \right) \, , \qquad 
\gamma - 1 \equiv - \frac{2}{\left( 2\omega + 3 \right) \left( 1 + \frac{1}{2\omega + 3} \right)} \,, \\
\beta - 1 &\equiv \frac{\Psi \frac{\partial \omega}{\partial \Psi}}{\left( 1 + \frac{1}{2\omega + 3} \right)^2} \left( \frac{1}{2\omega + 3} \right)^3 \,.
\end{align}
\end{subequations}
Here the expression for $\beta$ differs from the one presented in Ref.\ \citen{Erik}, because we inverted the normalization $\frac{\kappa^2}{8\pi}\frac{1}{\Psi} \left( 1 + \frac{1}{2\omega + 3} \right) \equiv 1$ in order to get rid of $\kappa^2$. This result also matches the early computation\cite{Berkin:1993bt} of $\gamma$ in the Jordan frame with constant $\mathcal{B}^{\mathfrak{J}}_{AB}$, as well as the general result for a single scalar field with a potential.\cite{Hohmann:2013rba}

One can show that if for Eqs.\ \eqref{PPN_in_Einstein_frame} we use the mapping \eqref{mapping_in_Einstein_frame} backwards and for Eqs.\ \eqref{PPN_in_Jordan_frame} we use the mapping \eqref{mapping_in_Jordan_frame} backwards then both generalize to
\begin{subequations}
	\label{PPN_in_generic_frame}
\begin{align}
G_{eff} &\equiv \frac{\kappa^2}{8\pi} \mathcal{I}_1  \left( 1 + \mathcal{I}_5 \right) \, , \qquad
\gamma - 1 \equiv -2 \left( \frac{ \mathcal{I}_5 }{ 1 + \mathcal{I}_5 } \right) \,, \\
\beta - 1 &\equiv \frac{ \left( \ln \mathcal{I}_1 \right)^{,A} \left( \ln \mathcal{I}_1 \right)^{,B} \left( \left( \ln \mathcal{I}_1 \right)_{,AB} - \frac{1}{2} \mathcal{F}_{AB,C} \left( \ln \mathcal{I}_1 \right)^{,C} \right) }{16 \left( 1 + \mathcal{I}_5 \right)^2} \,
\end{align}
\end{subequations}
where the indexes are raised with $\mathcal{F}^{AB}$. Hence, it is evident that physical observables are frame and parametrization independent since they transform as invariants.


\section{Summary}
\label{Summary}

We studied general multiscalar-tensor theories of gravity without derivative couplings. By introducing quantities that are invariant under a local Weyl rescaling of the spacetime metric and transform covariantly if the scalar fields are reparametrized, we generalized the formalism of the invariants that has been developed in the case of a single scalar field.\cite{JKSV_inv,Vilson} Just as in the latter, we were able to construct rather simple ``translation'' rules in the context of multiscalar-tensor theories of gravity as well. By invoking the prescription, one can neatly compare the results obtained in different frames and parametrizations by ``translating'' an expression under consideration to a generic frame. As an example we used the formalism to show that the results of the parametrized post-Newtonian approximation, calculated in the Einstein frame\cite{DamourEF} and in the Jordan frame,\cite{Erik,Berkin:1993bt} indeed are the same if expressed in a generic frame.

It would be interesting to see, how the formalism of invariant quantities generalizes and could help to explore the next generation of MSTG, namely the theories of multi-Galileons, multi-Horndeski and beyond \cite{Padilla:2012dx,Kobayashi:2013ina,Ohashi:2015fma}, where the conformal transformation of the metric seems to generalize to multi-disformal \cite{Watanabe:2015uqa}.


\section*{Acknowledgments}

This work was supported by the Estonian Research Council Grant No.\ IUT02-27 and by the European Union through the European Regional Development Fund (Project No.\ 3.2.0101.11-0029).




\begin{thebibliography}{00}


\bibitem{DamourEF}
T.\ Damour and G.\ Esposito-Far\`{e}se ``Tensor-multi-scalar theories of gravitation'' 
{\it Class.\ Quantum Grav.} \href{http://dx.doi.org/10.1088/0264-9381/9/9/015}{{\bf 9}, 2093} (1992).


\bibitem{Berkin:1993bt}
A.\ L.\ Berkin and R.\ W.\ Hellings ``Multiple field scalar - tensor theories of gravity and cosmology''
{\it Phys.\ Rev.\ D} \href{http://dx.doi.org/10.1103/PhysRevD.49.6442}{{\bf 49}, 6442} (1994),
%
arXiv:\href{http://arxiv.org/abs/gr-qc/9401033}{gr-qc/9401033}.


\bibitem{Kaiser:2010yu} 
D.\ I.\ Kaiser and A.\ T.\ Todhunter ``Primordial perturbations from multifield inflation with nonminimal couplings''
{\it Phys.\ Rev.\ D} \href{http://dx.doi.org/10.1103/PhysRevD.81.124037}{{\bf 81}, 124037} (2010), 
%
arXiv:\href{http://arxiv.org/abs/1004.3805}{1004.3805} [astro-ph.CO].


\bibitem{Kaiser:2012ak} 
D.\ I.\ Kaiser, E.\ A.\ Mazenc and E.\ I.\ Sfakianakis ``Primordial bispectrum from multifield inflation with nonminimal couplings''
{\it Phys.\ Rev.\ D} \href{http://dx.doi.org/10.1103/PhysRevD.87.064004}{{\bf 87}, 064004} (2013),
%
arXiv:\href{http://arxiv.org/abs/1210.7487}{1210.7487} [astro-ph.CO].


\bibitem{Greenwood:2012aj} 
R.\ N.\ Greenwood, D.\ I.\ Kaiser and E.\ I.\ Sfakianakis ``Multifield dynamics of Higgs inflation''
{\it Phys.\ Rev.\ D} \href{http://dx.doi.org/10.1103/PhysRevD.87.064021}{{\bf 87}, 064021} (2013),
%
arXiv:\href{http://arxiv.org/abs/1210.8190}{1210.8190} [hep-ph].


\bibitem{Kaiser:2013sna} 
D.\ I.\ Kaiser and E.\ I.\ Sfakianakis ``Multifield inflation after Planck: the case for nonminimal couplings''
{\it Phys.\ Rev.\ Lett.}  \href{http://dx.doi.org/10.1103/PhysRevLett.112.011302}{{\bf 112}, 011302} (2014),
%
arXiv:\href{http://arxiv.org/abs/1304.0363}{1304.0363} [astro-ph.CO].


\bibitem{Kallosh:2013daa} 
R.\ Kallosh and A.\ Linde ``Multi-field conformal cosmological attractors''
{\it J.\ Cosmol.\ Astropart.\ Phys.} \href{http://dx.doi.org/10.1088/1475-7516/2013/12/006}{{\bf 1312}, 006} (2013),
%
arXiv:\href{http://arxiv.org/abs/1309.2015}{1309.2015} [hep-th].


\bibitem{White:2012ya} 
J.\ White, M.\ Minamitsuji and M.\ Sasaki ``Curvature perturbation in multi-field inflation with non-minimal coupling''
{\it J.\ Cosmol.\ Astropart.\ Phys.} \href{http://dx.doi.org/10.1088/1475-7516/2012/07/039}{{\bf 1207}, 039} (2012),
%
arXiv:\href{http://arxiv.org/abs/1205.0656}{1205.0656} [astro-ph.CO].


\bibitem{White:2013ufa} 
J.\ White, M.\ Minamitsuji and M.\ Sasaki ``Non-linear curvature perturbation in multi-field inflation models with non-minimal coupling''
{\it J.\ Cosmol.\ Astropart.\ Phys.} \href{http://dx.doi.org/10.1088/1475-7516/2013/09/015}{{\bf 1309}, 015} (2013),
%
arXiv:\href{http://arxiv.org/abs/1306.6186}{1306.6186} [astro-ph.CO].


\bibitem{Watanabe:2015eia} 
Y.\ Watanabe and J.\ White ``Multifield formulation of gravitational particle production after inflation''
{\it Phys.\ Rev.\ D} \href{http://dx.doi.org/10.1103/PhysRevD.92.023504}{{\bf 92}, 023504} (2015),
%
arXiv:\href{http://arxiv.org/abs/1503.08430}{1503.08430} [astro-ph.CO].


\bibitem{Kuusk:2014sna} 
P.\ Kuusk, L.\ J\"arv and E.\ Randla ``Scalar-tensor and multiscalar-tensor gravity and cosmological models'' in {\it Algebra, Geometry and Mathematical Physics},
Springer Proc.\ Math.\ Stat., Vol.\ 85 (Springer, Berlin, 2014), p.\ \href{http://dx.doi.org/10.1007/978-3-642-55361-5_40}{661}.


\bibitem{Rinaldi:2013lsa} 
M.\ Rinaldi ``The dark aftermath of Higgs inflation''
{\it Eur.\ Phys.\ J.\ Plus} \href{http://dx.doi.org/10.1140/epjp/i2014-14056-8}{{\bf 129}, 56} (2014),
%
arXiv:\href{http://arxiv.org/abs/1309.7332}{1309.7332} [gr-qc].


\bibitem{Rinaldi:2014yta} 
M.\ Rinaldi ``Higgs dark energy''
{\it Class.\ Quantum Grav.}  \href{http://dx.doi.org/10.1088/0264-9381/32/4/045002}{{\bf 32}, 045002} (2015),
%
arXiv:\href{http://arxiv.org/abs/1404.0532}{1404.0532} [astro-ph.CO].


\bibitem{Rinaldi}
M.\ Rinaldi ``Dark energy as a fixed point of the Einstein Yang-Mills Higgs equations'' {\it J.\ Cosmol.\ Astropart.\ Phys.} \href{http://dx.doi.org/10.1088/1475-7516/2015/10/023}{{\bf 1510}, 023} (2015),
%
arXiv:\href{http://arxiv.org/abs/1508.04576}{1508.04576} [gr-qc].


\bibitem{Vardanyan:2015oha} 
V.\ Vardanyan and L.\ Amendola ``How can we tell whether dark energy is composed of multiple fields?''
{\it Phys.\ Rev.\ D} \href{http://dx.doi.org/10.1103/PhysRevD.92.024009}{{\bf 92}, 024009} (2015),
%
arXiv:\href{http://arxiv.org/abs/1502.05922}{1502.05922} [gr-qc].


\bibitem{Horbatsch:2015}
M.\ Horbatsch, H.\ O.\ Silva, D.\ Gerosa, P.\ Pani, E.\ Berti, L.\ Gualtieri and U.\ Sperhake ``Tensor-multi-scalar theories: relativistic stars and 3+1 decomposition'' {\it Class.\ Quantum Grav.}  \href{http://dx.doi.org/10.1088/0264-9381/32/20/204001}{{\bf 32}, 204001} (2015),  
%
arXiv:\href{http://arxiv.org/abs/1505.07462}{1505.07462} [gr-qc].


\bibitem{Steinwachs:2011zs} 
C.\ F.\ Steinwachs and A.\ Y.\ Kamenshchik ``One-loop divergences for gravity non-minimally coupled to a multiplet of scalar fields: calculation in the Jordan frame. I. The main results''
{\it Phys.\ Rev.\ D} \href{http://dx.doi.org/10.1103/PhysRevD.84.024026}{{\bf 84}, 024026} (2011),
%
arXiv:\href{http://arxiv.org/abs/1101.5047}{1101.5047} [gr-qc].


\bibitem{JKSV_inv}
L.\ J\"arv, P.\ Kuusk, M.\ Saal and O.\ Vilson ``Invariant quantities in the scalar-tensor theories of gravitation''
{\it Phys.\ Rev.\ D} \href{http://dx.doi.org/10.1103/PhysRevD.91.024041}{{\bf 91}, 024041} (2015),
%
arXiv:\href{http://arxiv.org/abs/1411.1947}{1411.1947} [gr-qc].


\bibitem{JKSV_trans}
L.\ J\"arv, P.\ Kuusk, M.\ Saal and O.\ Vilson ``Transformation properties and general relativity regime in scalar-tensor theories'' {\it Class.\ Quantum Grav.}  \href{http://dx.doi.org/10.1088/0264-9381/32/23/235013}{{\bf 32}, 235013} (2015),
%
arXiv:\href{http://arxiv.org/abs/1504.02686}{1504.02686} [gr-qc].


\bibitem{Vilson}
O. Vilson ``Some remarks concerning invariant quantities in scalar-tensor gravity'',
to appear in {\it Advances in Applied Clifford Algebras}, 
%
arXiv:\href{http://arxiv.org/abs/1509.02481}{1509.02481} [gr-qc].


\bibitem{Erik}
E. Randla ``PPN parameters for multiscalar-tensor gravity without a potential''
{\it J.\ Phys.: Conf.\ Ser.} \href{http://dx.doi.org/10.1088/1742-6596/532/1/012024}{{\bf 532}, 012024} (2014).


\bibitem{Flanagan}
\'E.\ \'E.\ Flanagan ``The conformal frame freedom in theories of gravitation'' 
{\it Class.\ Quantum Grav.} \href{http://dx.doi.org/10.1088/0264-9381/21/15/N02}{{\bf 21}, 3817} (2004),
%
arXiv:\href{http://arxiv.org/abs/gr-qc/0403063}{gr-qc/0403063}.


\bibitem{Bettoni_Zumalacarregui}
D. Bettoni and M. Zumalac\'arregui ``Kinetic mixing in scalar-tensor theories of gravity''
{\it Phys.\ Rev.\ D} \href{http://dx.doi.org/10.1103/PhysRevD.91.104009}{{\bf 91}, 104009} (2015),
%
arXiv:\href{http://arxiv.org/abs/1502.02666}{1502.02666} [gr-qc].


\bibitem{Bekenstein}
J.\ D.\ Bekenstein ``The relation between physical and gravitational geometry''
{\it Phys.\ Rev.\ D} \href{http://dx.doi.org/10.1103/PhysRevD.48.3641}{{\bf 48}, 3641} (1993),
%
arXiv:\href{http://arxiv.org/abs/gr-qc/9211017}{gr-qc/9211017}.


\bibitem{disformal}
M.\ Zumalac\'arregui and J.\ Garc\'{\i}a-Bellido ``Transforming gravity: from derivative couplings to matter to second-order scalar-tensor theories beyond the Horndeski Lagrangian''
{\it Phys.\ Rev.\ D} \href{http://dx.doi.org/10.1103/PhysRevD.89.064046}{{\bf 89}, 064046} (2014),
%
arXiv:\href{http://arxiv.org/abs/1308.4685}{1308.4685} [gr-qc].


\bibitem{Kaiser}
D.\ I.\ Kaiser ``Conformal transformations with multiple scalar fields''
{\it Phys.\ Rev.\ D} \href{http://dx.doi.org/10.1103/PhysRevD.81.084044}{{\bf 81}, 084044} (2010),
%
arXiv:\href{http://arxiv.org/abs/1003.1159}{1003.1159} [gr-qc].


\bibitem{Hohmann:2013rba} 
M.\ Hohmann, L.\ J\"arv, P.\ Kuusk and E.\ Randla ``Post-Newtonian parameters $\gamma$ and $\beta$ of scalar-tensor gravity with a general potential''
{\it Phys.\ Rev.\ D} \href{http://dx.doi.org/10.1103/PhysRevD.88.084054}{{\bf 88}, 084054} (2013) [Erratum: {\it ibid.} \href{http://dx.doi.org/10.1103/PhysRevD.89.069901}{{\bf 89}, 069901} (2014)],
%
arXiv:\href{http://arxiv.org/abs/1309.0031}{1309.0031} [gr-qc].


\bibitem{Padilla:2012dx} 
A.\ Padilla and V.\ Sivanesan ``Covariant multi-galileons and their generalisation''
{\it J.\ High Energy Phys.} \href{http://dx.doi.org/10.1007/JHEP04(2013)032}{{\bf 2013}, 32} (2013),
%
arXiv:\href{http://arxiv.org/abs/1210.4026}{1210.4026} [gr-qc].


\bibitem{Kobayashi:2013ina} 
T.\ Kobayashi, N.\ Tanahashi and M.\ Yamaguchi ``Multifield extension of $G$ inflation''
{\it Phys.\ Rev.\ D} \href{http://dx.doi.org/10.1103/PhysRevD.88.083504}{{\bf 88}, 083504} (2013),
%
arXiv:\href{http://arxiv.org/abs/1308.4798}{1308.4798} [hep-th].


\bibitem{Ohashi:2015fma} 
S.\ Ohashi, N.\ Tanahashi, T.\ Kobayashi and M.\ Yamaguchi ``The most general second-order field equations of bi-scalar-tensor theory in four dimensions''
{\it J.\ High Energy Phys.} \href{http://dx.doi.org/10.1007/JHEP07(2015)008}{{\bf 1507}, 008} (2015),
%
arXiv:\href{http://arxiv.org/abs/1505.06029}{1505.06029} [gr-qc].


\bibitem{Watanabe:2015uqa} 
Y.\ Watanabe, A.\ Naruko and M.\ Sasaki ``Multi-disformal invariance of non-linear primordial perturbations''
{\it Europhys.\ Lett.}  \href{http://dx.doi.org/10.1209/0295-5075/111/39002}{{\bf 111}, 39002} (2015),
%
arXiv:\href{http://arxiv.org/abs/1504.00672}{1504.00672} [gr-qc].

\end{thebibliography}
\end{document}